\documentclass{jetpl}
\twocolumn

\usepackage{caption}
\usepackage{graphicx}
\usepackage{cite}
\usepackage[dvips]{graphicx}
\DeclareGraphicsExtensions{.pdf,.png,.jpg,.eps,.ps}
\RequirePackage{caption}
\DeclareCaptionLabelSeparator{point}{. }
\title{Источники первичного космического излучения, формирующие бамп около $E_0~=~100~PeV$}
\rtitle{Источники первичного космического излучения, формирующие бамп\ldots}
\sodtitle{Источники первичного космического излучения, формирующие бамп\ldots}

\author{С.\,Е.\,Пятовский \/\thanks{ mail: vgsep@ya.ru\\
ORCID: 0000-0003-2565-1670}}
\rauthor{С.\,Е.\,Пятовский}
\sodauthor{Пятовский}

\address{Физический институт им.~П.~Н.~Лебедева Российской академии наук}

\dates{14 июля 2023}

\abstract
{Выполнен сравнительный анализ спектров первичного космического излучения (ПКИ) по $E_0$ и спектров переменных звезд по периодам, с целью установить причины нерегулярностей в спектре ПКИ по $E_0$. Показана зависимость между периодами переменных звезд и максимальной энергией $E_0$ ядер ПКИ, генерируемой данными типами звезд. Нерегулярности в спектре ПКИ по $E_0$ связаны с переходом с ростом $E_0$ от одного доминирующего типа звезд к другому. Излом в спектре ПКИ при $E_0~=~3-5~PeV$ связан с уменьшением вклада звезд переменности SRB и дальнейшим ростом вклада звезд переменности мириды в поток ПКИ. Бамп в спектре ПКИ с максимумом при $E_0~=~80~PeV$ образован звездами-гигантами и сверхгигантами переменностей мириды и SRC.}


\begin{document}

\maketitle
{\bf Введение.}
Причины и вид нерегулярностей спектра ПКИ по $E_0$ остается предметом научных дискуссий. Обсуждаются вопросы локализации т.н. изломов спектра, "острота" изломов, при каких энергиях наблюдаются изломы в спектрах легких и тяжелых ядер в массовом составе ПКИ и другие. Внимание вызывают вопросы локализации и источнике бампа в спектре ПКИ по $E_0$ около $100~PeV$.

\begin{figure}[h]
\centering
\includegraphics[width=\linewidth]{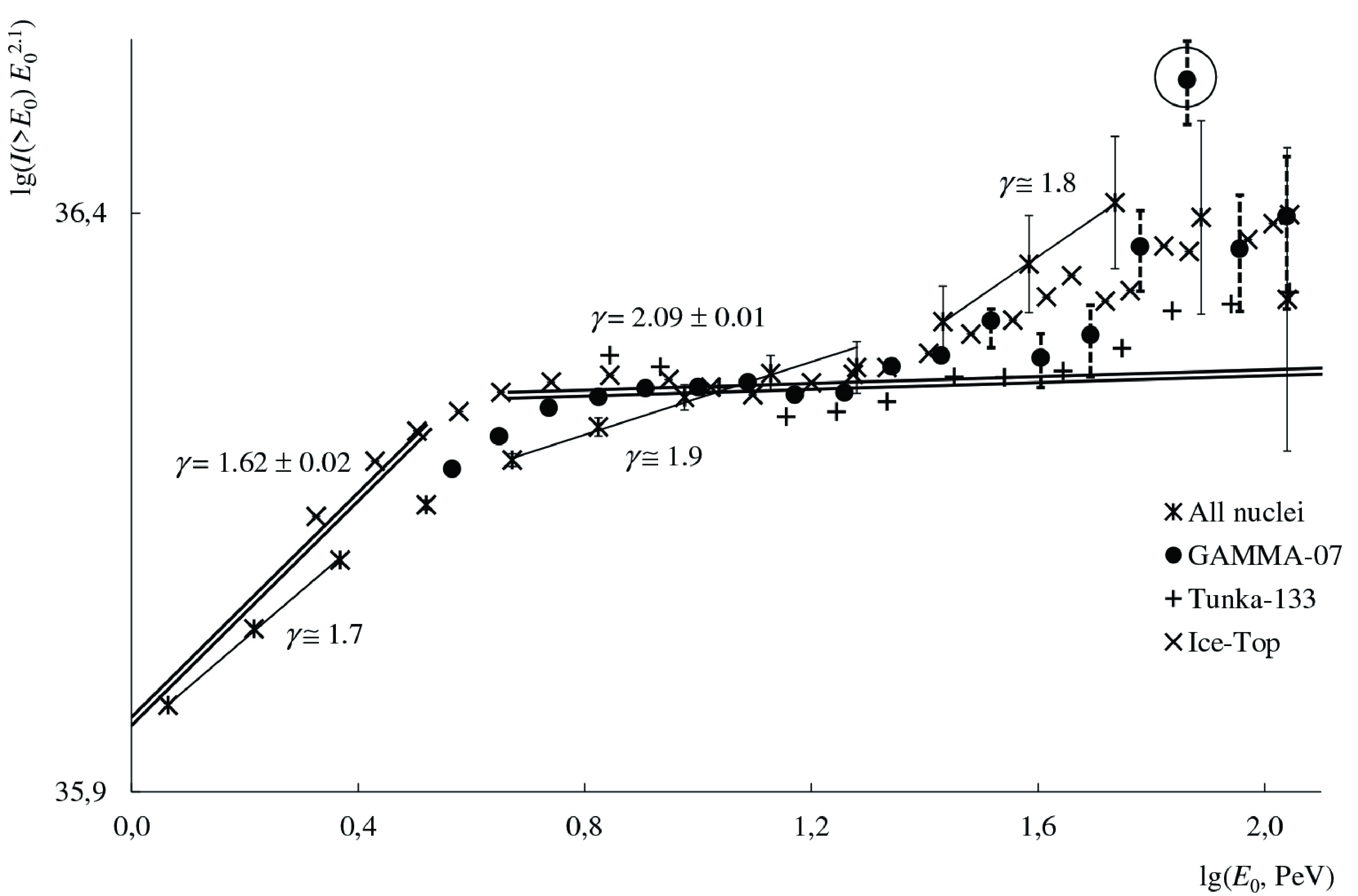}
\caption{Интегральные спектры по $E_0$, полученные в экспериментах \mbox{KASCADE-Grande} и \mbox{GAMMA-07}. Максимумы бампов в спектрах, полученных в экспериментах \mbox{KASCADE-Grande} и \mbox{GAMMA-07}, соответствуют $E_0~=~50-80~PeV$, что выше энергии излома самых тяжелых ядер в массовом составе ПКИ~\cite{1}. В спектре ПКИ по $E_0$, полученном в эксперименте \mbox{GAMMA-07}, наблюдается выпадающее значение (обведено).}
\label{fig1}
\end{figure}

Анализ нерегулярностей спектра ПКИ при $E_0~=~1-100~PeV$ выполнен, в частности, в работе~\cite{1}. На рисунке~\ref{fig1} показаны результаты экспериментов \mbox{KASCADE-Grande}~\cite{2}, Tunka и \mbox{Ice-Top} энергетического спектра ПКИ. Отдельного внимания заслуживают результаты экспериментов GAMMA (\mbox{GAMMA-07})~\cite{3} (Армения, Арагац) и "Адрон" (Тянь-Шаньская высокогорная научная станция), в которых при $E_0~\cong~70-100~PeV$ зарегистрирован пик интенсивности, существенно превышающий данные других экспериментов. Природа данной нерегулярности не установлена при том, что другие эксперименты о наличии аналогичного пика не сообщают. Необычным является и то, что данный пик не наблюдался в экспозициях самого эксперимента GAMMA за другие периоды времени, например, \mbox{GAMMA-06}, -08 и др. Однако вполне возможно, что данный пик не является методической погрешностью обработки экспериментальных данных.

В работах~\cite{1,4} показано, что нерегулярности в спектре ПКИ по $E_0$, следующего за изломом при $E_0~=~3-5~PeV$, обусловлены выбытием ядер массового состава ПКИ начиная с протонов. Методом "мини-макс возраста ШАЛ"~\cite{1}, основанным на существенно большой статистике экспериментальных характеристик ШАЛ, полученных, в частности, в эксперименте \mbox{KASCADE-Grande}, показано, что при $E_0~=~2-35~PeV$ массовый состав ядер ПКИ остается смешанным и соответствующим CNO-группе. Однако излом в спектре ядер массового состава ПКИ самой тяжелой группы локализован до бампа, наблюдаемого при $E_0~=~50-100~PeV$, что указывает на то, что бамп в спектре ПКИ при $E_0~=~50-100~PeV$ образован другими источниками ядер с другими особенностями ускорения.

{\bf 1 Экспериментальные данные для анализа бампа при \boldmath$E_0~=~50-100~PeV$.}
Анализ бампа спектра ПКИ по $E_0$ выполнен по данным эксперимента \mbox{KASCADE-Grande}~\cite{2,4}, база данных которого содержит характеристики более 150 млн ШАЛ, в т.ч. глобальное время регистрации ШАЛ.

Характеристикой данной нерегулярности (бампа) является показатель наклона $\gamma$ спектра ПКИ по $E_0$. Для оценки изменения $\gamma$ выбран диапазон по $E_0~=~20-75~PeV$, расположенный после излома группы самых тяжелых ядер в массовом составе ПКИ и до максимума бампа при $E_0~=~80~PeV$, подтвержденного в экспериментах GAMMA и "Адрон". Изучение изменения $\gamma$ выполнено с лагом 10 дней, что обеспечило статистику выборок $\cong$~1~млн событий.

\begin{figure}[h]
\centering
\includegraphics[width=\linewidth]{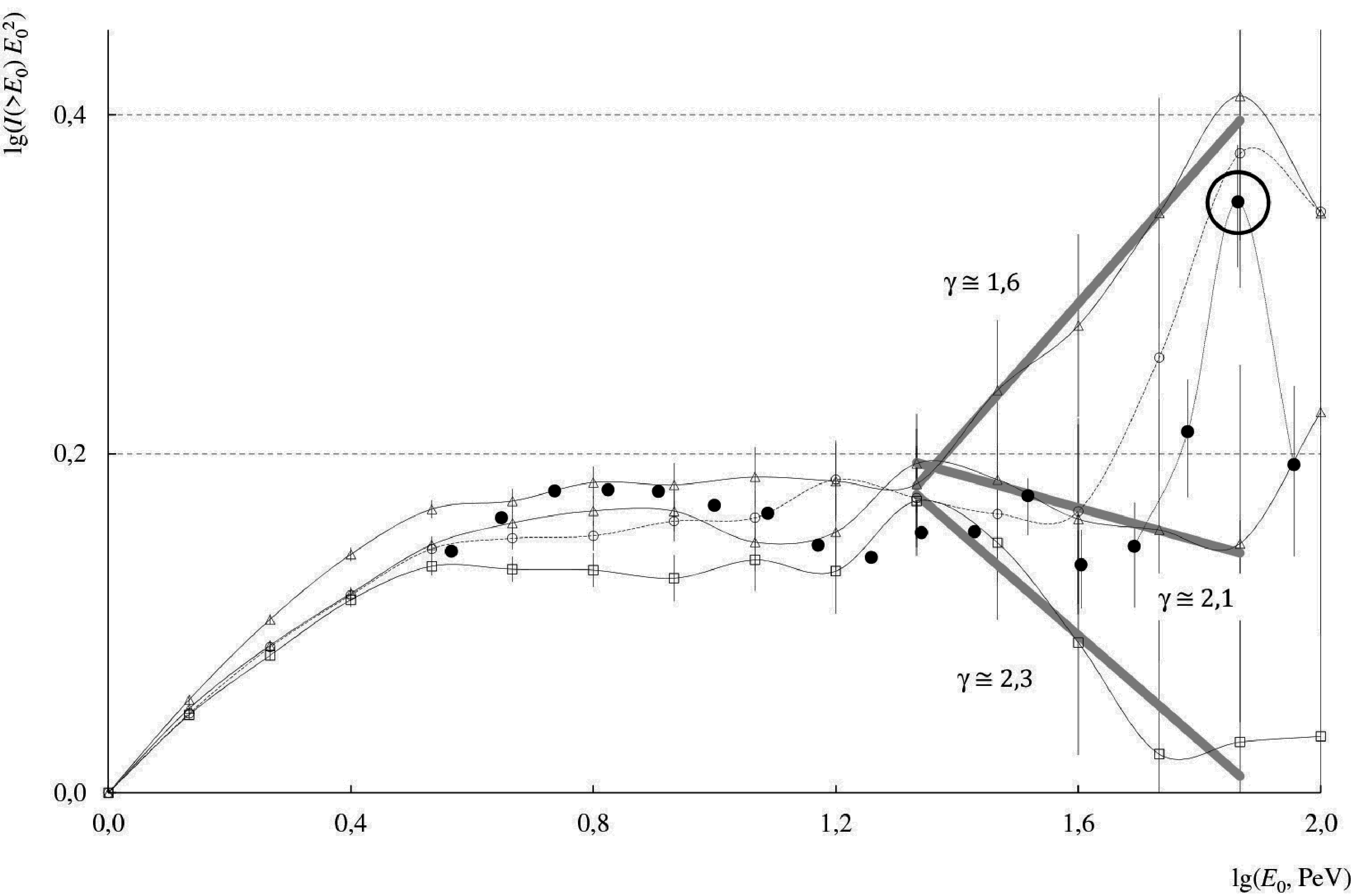}
\caption{Спектр ПКИ по $E_0$, полученный в эксперименте \mbox{GAMMA-07} (черные кружки, выпадающее значение обведено), в сравнении с данными эксперимента \mbox{KASCADE-Grande} для различных интервалов времени (пустые маркеры). Серые прямые – регрессии в интервале $E_0~=~20-75~PeV$.}
\label{fig2}
\end{figure}

Примеры полученных спектров приведены на рисунке~\ref{fig2}, где показаны спектры с показателями наклона $\gamma$ вблизи $E_0~=~80~PeV$ от минимальных значений $\gamma~=~1.60~\pm~0.02$ до максимальных $\gamma~=~2.31~\pm~0.04$. Спектры, построенные по выборкам из экспериментальных данных KASCADE-Grande, приведены в сравнении с данными эксперимента \mbox{GAMMA-07}. Несмотря на то, что усредненный по всей статистике наблюдений показатель $\gamma$ получен с высокой точностью, значения $\gamma$ за различные интервалы времени существенно различаются. Данное изменение $\gamma$ может быть связано либо с флуктуациями характеристик ШАЛ, либо с изменением интенсивности ПКИ в данном диапазоне $E_0$. Из рисунка~\ref{fig2} следует, что выпадающее событие, зарегистрированное в эксперименте \mbox{GAMMA-07}, не является уникальным и имеет аналоги в событиях, зарегистрированных в эксперименте \mbox{KASCADE-Grande}.

По базе данных эксперимента \mbox{KASCADE-Grande} получены значения показателя $\gamma$ в диапазоне по $E_0~=~20-75~PeV$ для 248 временных интервалов.

\begin{figure}[h]
\centering
\includegraphics[width=\linewidth]{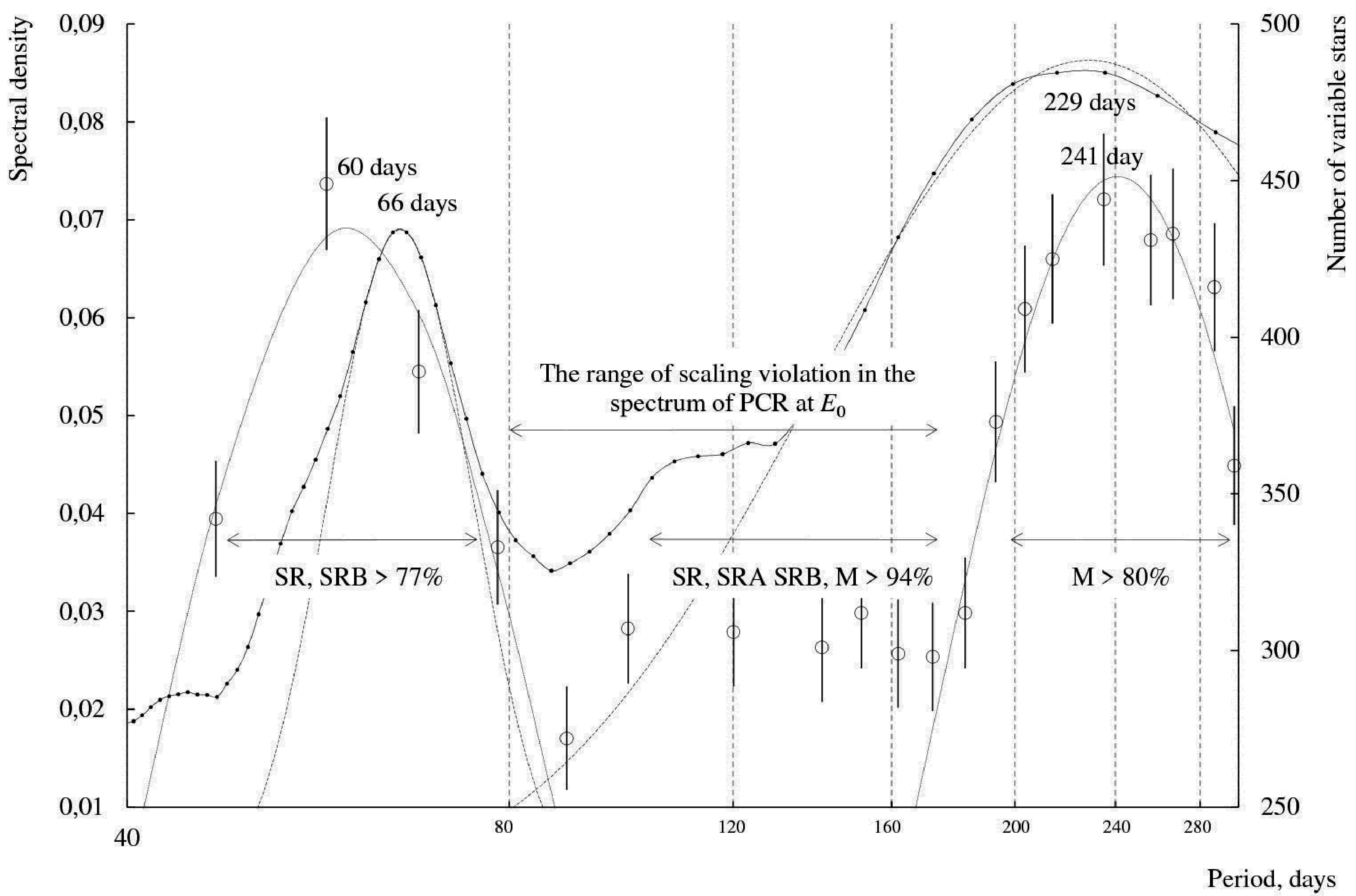}
\caption{Спектральная плотность лог-периода изменения показателя $\gamma$ спектра ПКИ по $E_0$ (сплошная линия, левая ось) в сравнении с периодами звезд различных типов переменности (пунктирные линии, правая ось): SR – полурегулярные красные гиганты и сверхгиганты промежуточных или поздних спектральных классов, SRA и SRB – полурегулярные красные гиганты поздних спектральных классов М, С и S, М – мириды, красные гиганты на конечных этапах звездной эволюции с эмиссионными спектрами поздних классов.}
\label{fig4}
\end{figure}

{\bf 2 Спектральный анализ изменения показателя \boldmath$\gamma$.}
Спектральный анализ изменения $\gamma$ выполнен с целью выявления возможных максимумов периодов изменения значений $\gamma$. Для анализа применено спектральное преобразование Фурье с окном Хэмминга. Полученная спектральная плотность лог-периода приведена на рисунке~\ref{fig4}. Проведенный анализ позволил выявить на интервале $40-300$ дней два максимума в периоде изменения $\gamma$ (66 и 229 дней). Ширина пика спектральной плотности характеризует "локальность" источника ПКИ, – чем ближе пик к нормальному распределению, тем с большей вероятностью в формировании пика доминирует один источник ПКИ. На рисунке~\ref{fig4} пики с максимумами в периодах 66 и 229 дней описываются нормальными распределениями с $R_a^2~>~98\%$.

Для поиска возможных источников ПКИ в диапазоне $E_0~=~20-100~PeV$ использованы каталоги звездных объектов "General Catalogue of Variable Stars (GCVS)"~\cite{5} и "Zwicky Transient Facility Catalog (ZTF)"~\cite{6}. В GCVS представлено более 60 тыс. звезд более 250 типов с указанием периодов, локаций и других характеристик. На рисунке~\ref{fig4} показано, что первой гармонике 66 дней соответствуют, в основном, звезды с переменностью типа SR, вторая гармоника 229 дней образована преимущественно миридами. Здесь же необходимо обратить внимание, что звезды, находящиеся на заключительных этапах эволюции, обычно имеют сильные магнитные поля.

Область перехода от полурегулярных гигантов к миридам (рисунок~\ref{fig4}) характеризуется локальным нарушением скейлинга в спектре ПКИ при $E_0~=~3-20~PeV$~\cite{7}. Локальных областей, аналогичных показанным на рисунке~\ref{fig4}, где происходит нарушение скейлинга, в спектре ПКИ по $E_0$ несколько, – области нарушения скейлинга связаны с переходом от одного доминирующего типа звезд к другому, а степени проявления нарушений скейлинга определены распределениями по энергиям, которые обеспечены доминирующим типом звезд рассматриваемой переменности.

\begin{figure*}[h]
\centering
\includegraphics[width=\linewidth]{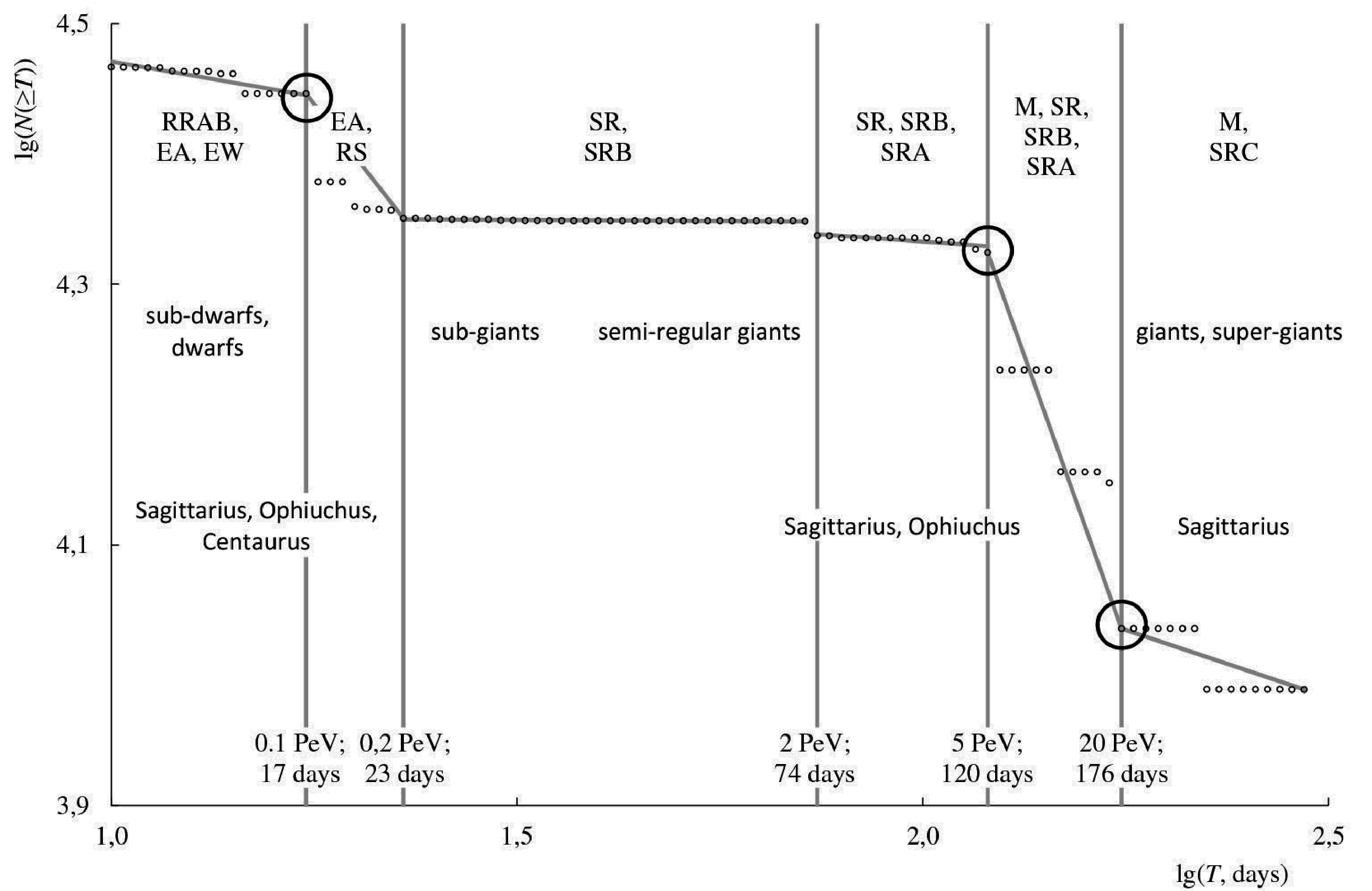}
\caption{Интегральный спектр переменных звезд по лог-периоду. Обозначены доминирующие типы звезд и созвездия, где находятся данные звезды, для указанных интервалов по $E_0$ и периодам: EA – оранжевые субгиганты поздней эволюции, EW – контактные желтые карлики спектрального класса F, RS – эруптивные желто-белые карлики со вторичным компонентом магнитно-активным субгигантом со спектрами Ca-II, H и K в эмиссии, SRC – полурегулярные сверхгиганты поздних спектральных классов М, С и S. Обведены нерегулярности в спектре с соответствующими значениями $E_0$.}
\label{fig5}
\end{figure*}

{\bf 3 Спектр звезд по периоду.}
Интегральный спектр переменных звезд в зависимости от лог-периода представлен на рисунке~\ref{fig5}, где показаны усредненные по типам переменных звезд периоды. Рассмотрены звезды типов от белых карликов до сверхгигантов типа рекуррентных новых. Чем более существенны нерегулярности в спектре по периоду источников ПКИ, тем более существенны нерегулярности в спектре ПКИ по $E_0$, образованном данными источниками. Самые большие нерегулярности в спектре по периоду обозначены на рисунке~\ref{fig5} как известные значения $E_0$: периоду 17 дней соответствует $E_0~=~0.1~PeV$ (область красных карликов), 120 дней, – $E_0~=~5~PeV$ (бамп при $E_0~=~3-5~PeV$ в спектре ПКИ) и 176 дней, – $E_0~=~20~PeV$ (начало бампа с максимумом при $E_0~=~80~PeV$). Ускорение ПКИ до $E_0~=~0.1~PeV$ во вспышках красных карликов установлено в работах Ю.~И.~Стожкова~\cite{8}.

На рисунке~\ref{fig5} показано, что источниками ПКИ низких энергий $E_0~<~0.1~PeV$ являются карлики, локализованные преимущественно в созвездиях Стрельца, Змееносца и Центавра, средних энергий $E_0~=~0.2-2~PeV$, – субгиганты и гиганты из созвездий Стрельца и Змееносца, и высоких энергий $E_0~>~5~PeV$, – гиганты и сверхгиганты из созвездия Стрельца.

Из анализа данных, представленных на рисунке~\ref{fig5}, получена регрессия, определяющая зависимость между средним периодом данного типа звезды-источника ПКИ и максимальной $E_0$:

\begin{multline}
\mathrm{lg}(T, \mathrm{days})~=~(0.45~\pm~0.05)\mathrm{lg}(E_0, \mathrm{PeV})~+\\
+~(1.71~\pm~0.05)
\label{multline:1}
\end{multline}

Из регрессии (1) следует, что нижняя граница области субгигантов для периода 23 дня составляет $E_0~=~0.2~PeV$, область начала первого бампа в спектре ПКИ по $E_0$, – период 74 дня и $E_0~=~2-3~PeV$. Также можно оценить максимальную энергию ПКИ: согласно каталогу GCVS~\cite{5}, максимальный зарегистрированный период составляет 29000 дней (80 лет) для рекуррентных новых звезд переменности типа NR, что соответствует максимально зарегистрированной в КЛ $E_0~\cong~1-2~ZeV$.

{\bf 4 Типы переменных звезд и спектр ПКИ по \boldmath$E_0$.}
Из рисунков~\ref{fig4} и~\ref{fig5} следует, что полурегулярные гиганты и мириды составляют основное звездонаселение, обеспечивающее источники ПКИ при $E_0~=~1-100~PeV$. С применением метода основного массива рассмотрим формирование звездами переменностей типа SR, SRA, SRB и М спектра ПКИ при данных $E_0$.

Распределение звезд по лог-периодам соответствует нормальному распределению $N~\sim~\mathrm{exp}(-\frac{(\mathrm{ln}(T)-\mathrm{ln}(\bar{T}))^2}{2\sigma^2})$. Наиболее близкими к периоду 229 дней, полученному Фурье-анализом по изменению показателя $\gamma$ спектра ПКИ по $E_0$ (рисунок~\ref{fig4}), становятся звезды переменностей SRA (193 дня), М (280 дней) и SRC (372 дня). С учетом, что количество мирид существенно больше, нежели звезд других типов с близкими периодами, можно предположить, что бамп в спектре ПКИ по $E_0$ около $100~PeV$ образован, в основном, миридами. С ростом $E_0$ в диапазоне $1-100~PeV$ определяющий вклад в ПКИ начинают вносить звезды на конечных этапах звездной эволюции, что вызывает утяжеление массового состава ПКИ. Однако для каждого значения $E_0$ массовый состав ПКИ определяется доминирующим типом звезд данного периода, что может приводить к существенным флуктуациям доли различных ядер в массовом составе ПКИ с изменением по $E_0$.

\begin{figure}[h]
\centering
\includegraphics[width=\linewidth]{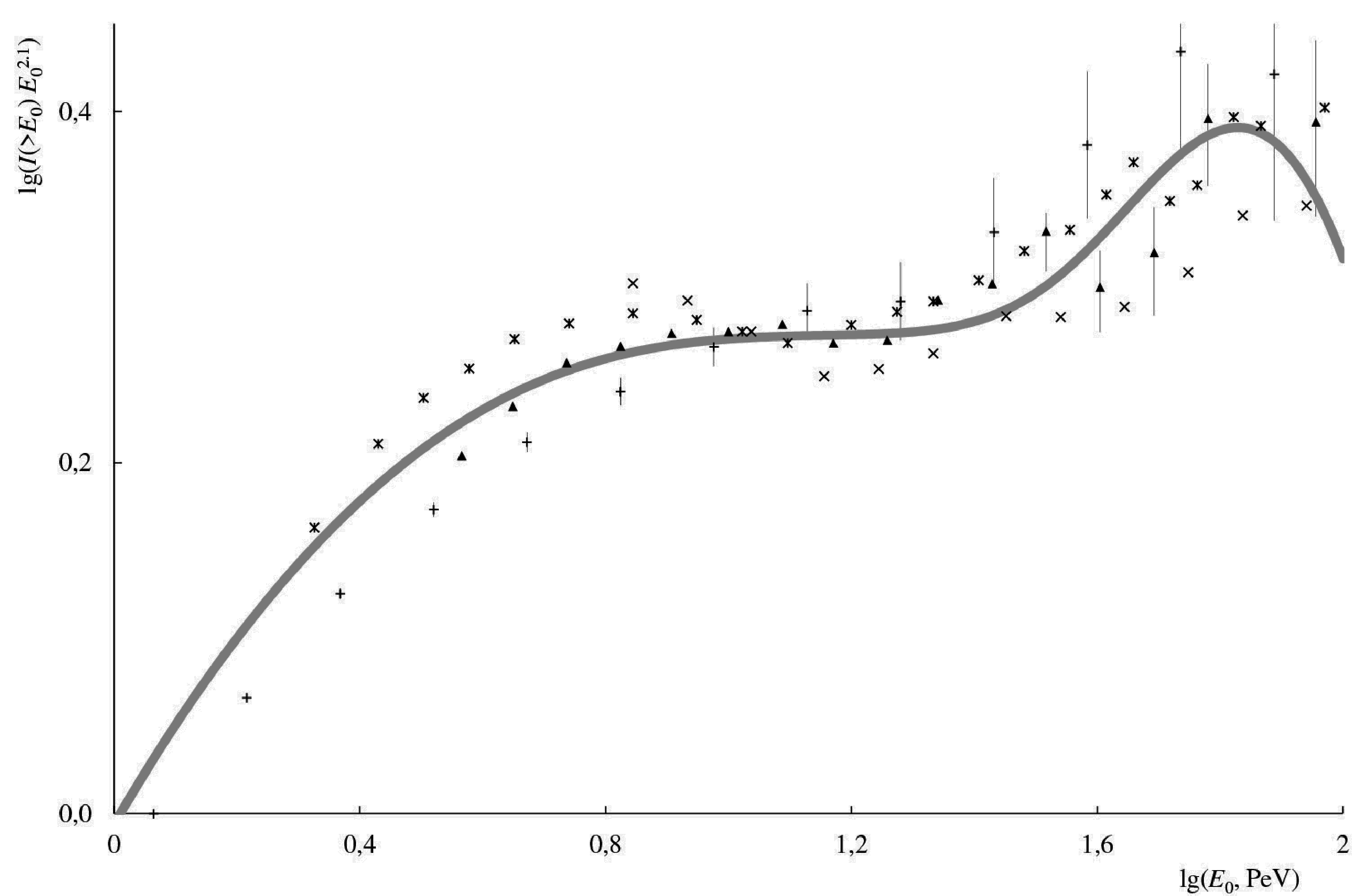}
\caption{Сравнение спектров ПКИ по $E_0$, полученных в экспериментах GAMMA, \mbox{Tunka-133}, \mbox{Ice-Top} и \mbox{KASCADE-Grande} с распределением звезд переменностей типа SRA, SRB, SRC и М (серая кривая).}
\label{fig8}
\end{figure}

Спектры ПКИ по $E_0$, полученные в экспериментах GAMMA, \mbox{Tunka-133}, \mbox{Ice-Top} и \mbox{KASCADE-Grande}, в сравнении со спектрами доминирующих звезд переменностей SR, SRA, SRB, SRC и М показаны на рисунке~\ref{fig8}. Максимум бампа спектра ПКИ в данном случае составляет $E_0~\cong~67~PeV$.

Однако, как следует из рисунка~\ref{fig8}, бамп должен быть менее выраженным и находиться при $E_0~<~67~PeV$. Средний период звезд переменности SRC (сверхгиганты) 372 дня, что по (1) дает значение lg$(E_0)~=~1.91 (81~PeV)$. Данное значение $E_0$ соответствует результатам эксперимента \mbox{KASCADE-Grande}. В то же время средний период мирид 280 дней или lg$(E_0)~=~1.64 (44~PeV)$. Т.к. количество наблюдаемых мирид на порядок больше, нежели звезд переменности SRC, локализация максимума бампа по данным эксперимента KASCADE-Grande при $E_0~=~80~PeV$ завышена, и бамп сформирован как миридами, так и звездами-сверхгигантами переменности SRC.

Ускорение КЛ до сверхвысоких энергий происходит во взрывных и новоподобных звездах, например, в рекуррентных новых звездах, у которых зарегистрирован максимальный период, обеспечивающий максимальную $E_0~\sim~1-2~ZeV$. В интервале $E_0~=~200~PeV~-~3.5~EeV$ типу переменности ZAND соответствует средний период $T~=~553$, или по~(1) $E_0~=~200~PeV$, далее следуют звезды переменности $N$ с $T~=~2000$, или $E_0~=~3.5~EeV$, что обеспечивает минимум в спектре ПКИ при данных $E_0$.

Примером двойной звездной системы, где происходит ускорение до сверхвысоких энергий, может быть звезда переменности EA+SRC типа, с зарегистрированным периодом 7430 дней (20 лет), что должно обеспечить бамп в спектре ПКИ при $E_0~=~60~EeV$ или lg$(E_0,~PeV)~=~4.80$. Это может быть звезда типа $\mu$~Цефея (гранат Гершеля, красный сверхгигант на последней стадии звездной эволюции с Не-С циклом, что указывает на то, что массовый состав ПКИ относительно CNO-группы становится более легким при данных $E_0$) и звезда типа Алголь $\beta$~Персея.

Примером тройной звездной системы, где происходит ускорение до сверхвысоких энергий, может быть система звезд с зарегистрированным периодом 11900 дней (33 года), что должно обеспечить бамп в спектре ПКИ при $E_0~=~180~EeV$ или lg$(E_0,~PeV)~=~5.26$. Если данный источник считать единственным, обеспечивающим поток ПКИ при $E_0~=~180~EeV$, изменения потока ПКИ должны быть существенны, – от максимума до полного затухания, что наблюдается в экспериментах.

\begin{figure}[h]
\centering
\includegraphics[width=\linewidth]{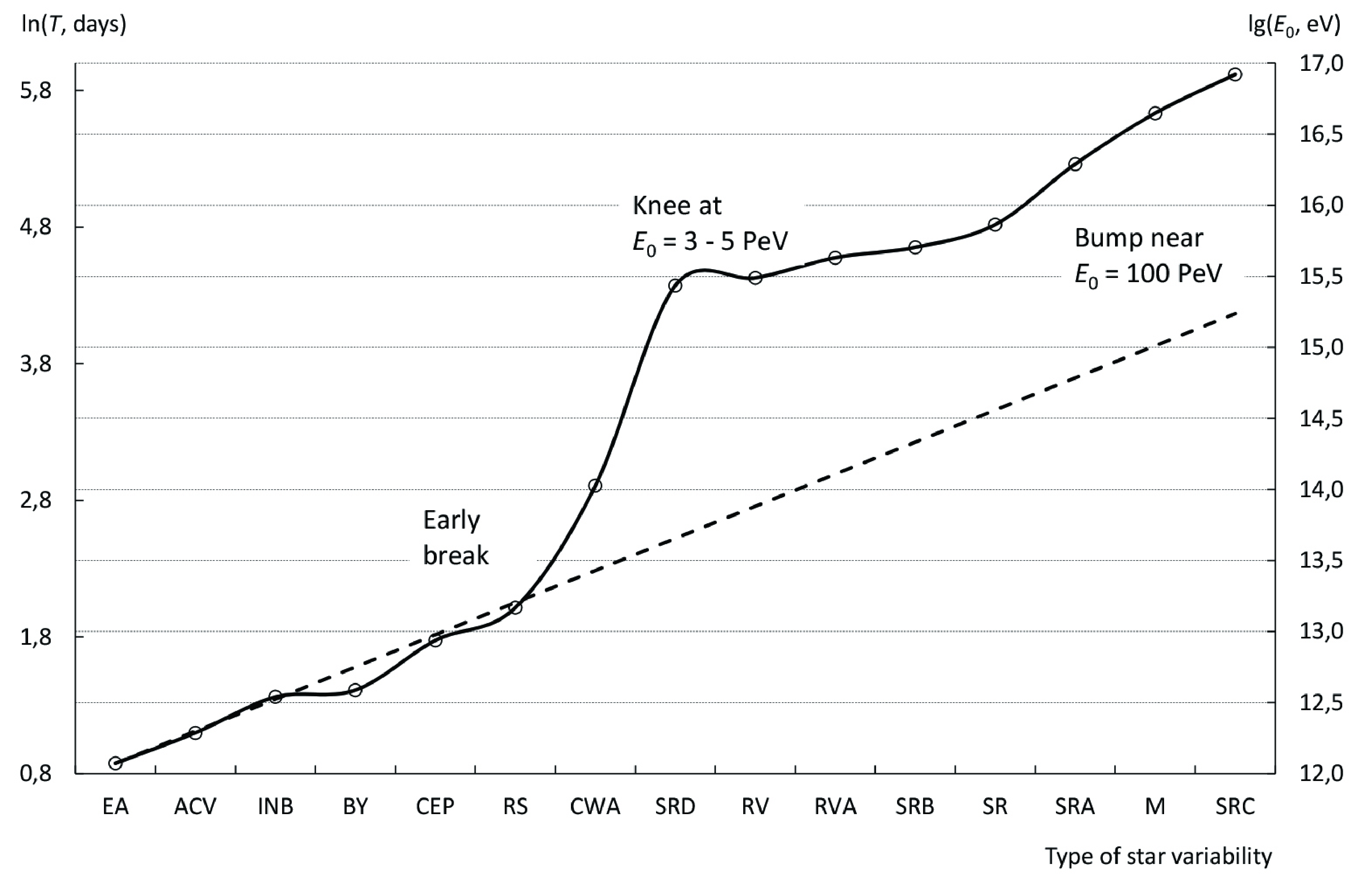}
\caption{Зависимость периодов звезд и энергии $E_0$ от типов затменно-переменных звезд.}
\label{fig10}
\end{figure}

Суммируя полученные в данном исследовании результаты, построим зависимость лог-периода и $E_0$ от типов затменно-переменных звезд, приведенную на рисунке~\ref{fig10}. Данный спектр характеризуется тремя основными областями нерегулярностей относительно линейной составляющей: начиная со звезд переменности RS наблюдается "ранний" излом в спектре ПКИ по $E_0$; начиная со звезд типа SRD (гиганты и сверхгиганты спектральных классов F, G, K), – излом при $E_0~=~3-5~PeV$; начиная с мирид, – т.н. "бамп" при $E_0$ около $100~PeV$. Как следует из рисунка~\ref{fig10}, после укручения спектра ПКИ по $E_0$ после излома при $3-5~PeV$, показатель $\gamma$ спектра ПКИ по $E_0$ вновь уменьшается и становится примерно таким же, каким был до излома.

{\bf Выводы.}
\begin{enumerate}
\item Источниками ПКИ являются переменные звезды различных типов, находящиеся на различных этапах эволюции, от субкарликов до сверхгигантов.
\item Существует зависимость между средним периодом для звезды данного типа переменности и максимальной энергией $E_0$ ПКИ, обеспечиваемой механизмами ускорения в данных звездах.
\item Звезда каждого типа своими вспышками определяет диапазон по $E_0$ ПКИ. Каждому диапазону по $E_0$ соответствует массовый состав ПКИ, определяемый типом звезды-источника и который меняется при изменении $E_0$.
\item Бамп в спектре ПКИ при $E_0$ около $100~PeV$ образован гигантами и сверхгигантами переменности M и SRC поздних спектральных классов. За другие нерегулярности в спектре ПКИ по $E_0$ ответственны другие типы звезд.

\end{enumerate}


\begin{thebibliography}{99}
\bibitem{1}
Erlykin~A.~D., Puchkov~V.~S., Pyatovsky~S.~E. Change in the mass composition of primary cosmic radiation at energies in the range of $E_0~=~1-100~PeV$ according to data of the \mbox{KASCADE-Grande} experiment//Physics of Atomic Nuclei. - 2021. - Vol. 84. - No 3. - p. 279-286. - DOI: 10.1134/S1063778821030170

\bibitem{2}
T.~Antoni, W.~D.~Apel, F.~Badea, K.~Bekk, A.~Bercuci, H.~Blumer, H.~Bozdog, I.~M.~Brancus, C.~Buttner, A.~Chilingarian, K.~Daumiller, P.~Doll, J.~Engler, F.~Febler, H.~J.~Gils, R.~Glasstetter, et al., Nucl. Instrum. Methods Phys. Res., Sect. A 513, 490 (2003). - DOI: 10.1016/S0168-9002(03)02076-X

\bibitem{3}
A.~P.~Garyaka, R.~M.~Martirosov, S.~V.~Ter-Antonyan, A.~D.~Erlykin, N.~M.~Nikolskaya, Y.~A.~Gallant, L.~W.~Jones, and J.~Procureur, J. Phys. G: Nucl. Part. Phys. 35, 115201 (2008); arXiv: 0808.1421v1 [astro-ph]. - DOI: 10.1088/0954-3899/35/11/115201

\bibitem{4}
Apel~W., Arteaga~J.~C., et al. The \mbox{KASCADE-Grande} experiment//\mbox{KASCADE-Grande} Collaborations, Nucl. Instrum. Methods Phys. Res. A 620 (2-3) (2010), pp. 202-216. - DOI: 10.1016/j.nima.2010.03.147

\bibitem{5}
General Catalogue of Variable Stars//The Sternberg Astronomical Institute, The Institute of Astronomy of Russian Academy of Sciences. URL: http://www.sai.msu.su/gcvs/

\bibitem{6}
C.~Xiaodian, W.~Shu, D.~Licai, et al. The Zwicky Transient Facility Catalog of Periodic Variable Stars//The Astrophysical Journal Supplement Series, 249:18 (21pp), 2020 July. DOI - 10.3847/1538-4365/ab9cae

\bibitem{7}
S.~B.~Shaulov, V.~A.~Ryabov, A.~L.~Schepetov, S.~E.~Pyatovsky, et al. Strange quark matter and the astrophysical nature of anomalous effects in cosmic rays at energies of $1-100~PeV$//Letters to the Journal of Experimental and Theoretical Physics. 2022. 1-2(7). 116. с. 3-12. DOI - 10.31857/S1234567822130018

\bibitem{8}
V.~G.~Sinitsyna, V.~Yu.~Sinitsyna, Yu.~I.~Stozhkov. Red dwarf stars as a new source type of galactic cosmic rays//Astronomische Nachrichten. 2021. 342. 1-2. pp. 342-346. DOI - 10.1002/asna.202113931

\end{thebibliography}
\end{document}